\documentclass[a4paper,twoside]{article}

\usepackage{epsfig}
\usepackage{subcaption}
\usepackage{calc}
\usepackage{amssymb}
\usepackage{amstext}
\usepackage{amsmath}
\usepackage{amsthm}
\usepackage{multicol}
\usepackage{pslatex}
\usepackage{apalike}
\usepackage{graphics}
\usepackage{SCITEPRESS}     

\begin{document}

\title{Greedy Scheduling: A Neural Network Method to Reduce Task Failure in Software Crowdsourcing}


\author{\authorname{Jordan Urbaczek
\sup{1}, Razieh Saremi \sup{1}, Mostaan Lotfalian Saremi\sup{1}
 and Julian Togelius \sup{2}}
\affiliation{\sup{1}School of Systems and Enterprises, Stevens Institute of Technology, Hoboken, NJ, USA}
\affiliation{\sup{2}Tandon School of Engineering,New York University, NYC, NY,USA}
\email{\{rsaremi, jurbacze,  mlotfali\}@stevens.edu, julian.togelius@nyu.edu}
}


\abstract{Highly dynamic and competitive crowdsourcing software development (CSD) marketplaces may experience task failure due to unforeseen reasons, such as increased competition over shared supplier resources, or uncertainty associated with a dynamic worker supply. Existing analysis reveals an average task failure ratio of 15.7\% in software crowdsourcing markets.These lead to an increasing need for scheduling support for CSD managers to improve the efficiency and predictability of crowdsourcing processes and outcomes.
To that end, this research proposes a task scheduling method based on neural networks, and develop a system that can predict and analyze task failure probability upon arrival. More  specifically, the model uses a range of input variables, including the number  of  open  tasks in  the  platform, the average  task  similarity  between arriving tasks and open tasks, the winner's monetary prize, and task duration, to predict  the  probability  of  task  failure on the planned arrival date and two surplus days. This prediction will offer the recommended day associated with lowest task failure probability to post the  task. The model on average provided 4\% lower failure probability per project. 
The proposed model empowers crowdsourcing managers to explore potential crowdsourcing outcomes with respect to different task arrival strategies.}

\onecolumn 
\maketitle

\setcounter{footnote}{0}
\section{\uppercase{Introduction}}
\label{sec:introduction}

Crowdsourced Software Development (CSD) has been used increasingly to develop software applications\cite{stol2014two} \cite{stol2014two}.  Crowdsourcing mini software development tasks leads to lower accelerated development \cite{saremi2017leveraging}. In order for a CSD platform to function efficiently, it must address both the needs of task providers as demands and crowd workers as suppliers. Any kind of skew in addressing these needs leads to task failure in the CSD platform. Generally, planning for CSD tasks that are complex, independent, and require a significant amount of time, effort, and expertise \cite{stol2014two} is challenging. For the task provider, requesting a crowdsourcing service is even  more challenging due to the uncertainty of the similarity among available tasks in the platform and the arrival of new tasks \cite{difallah2016scheduling}\cite{saremi2018hybrid}\cite{saremi2019ant}. The availability of crowd workers’ skill sets and consistency of performance history is also uncertain \cite{karim2016decision}\cite{zaharia2010delay}. These factors raise the issue of receiving qualified submissions, since crowd workers may be interested in multiple tasks from different task providers based on their individual utility factors \cite{faradani2011s}. 

It has been reported that crowd workers are more interested in working on tasks with similar concepts, monetary prize, technologies, complexities, priorities, and duration \cite{gordon1961general}\cite{faradani2011s}\cite{yang2015award}\cite{difallah2016scheduling}\cite{mejorado2020study}. However, attracting workers to a large group of similar tasks may cause zero registration, zero submissions, or unqualified submissions for some tasks due to lack of availability from workers\cite{khanfor2017failure}\cite{khazankin2011qos}. Moreover, lower level of task similarity in the platform leads to higher chance of task success and workers’ elasticity\cite{10.1007/978-3-030-50017-7_7}. 

For example, in Topcoder
a well-known Crowdsourcing Software platform, an average of 13 tasks arrive daily and are added to an average list of 200 existing tasks.  There is an average of 137 active workers to take the tasks at that period, which leads to an average of 25 failed tasks each day. According to this example, there will be a long queue of tasks waiting to be taken. Considering the fixed submission date, such a waiting line may result in failed tasks. These challenges have traditionally been addressed with task scheduling methods. 

The objective of this study is to provide a task schedule recommendation framework for a software  crowdsourcing platform  in  order  to  improve  the  success  and  efficiency  of software crowdsourcing. In this study, we first present a motivational example to explain the current task status in a software crowdsourcing platform. Then we propose a task scheduling architecture using a neural network strategy to reduce probability of task failure in the platform.

More  specifically, the system uses a range of input variables, including the number  of  open  tasks in  the  platform, the average  task  similarity  between arriving tasks and open tasks, the winner's monetary prize, and task duration, to predict  the  probability  of  task  failure on the planned arrival date and two surplus days. This prediction will offer the recommended the day associated with lowest task failure probability to post the  task.
The proposed system represents a task scheduling method for competitive crowdsourcing platforms based on the workflow of Topcoder, one of the primary software crowdsourcing platforms.  The   evaluation results provided on average 4\% lower task failure probability.

The remainder of this paper is structured as follows. Section II introduces a motivational  example that inspires this study. Section III presents background and review of available works. Section IV outlines our research design and methodology. Section V presents the case study and model evaluation, and Section VI presents the conclusion and outlines a number of directions for future work.

\section{\uppercase{Motivating Example}}

The motivation example illustrates a real crowdsourcing software development (CSD) project on the TopCoder platform. It was comprised of 41 tasks with a total project duration of 207 days, with an average of 8 days per task.
The project experienced a 47\% task failure ratio, which means 19 of the 41 tasks failed. 6 tasks failed due to client requests ( i.e  14\% failure) and 7 tasks failed due to failed requirements (17\% failure). The remaining 8 tasks (i.e 14\% failure) failed due to zero submissions.

\begin{figure}[ht!]
\includegraphics[width=1\columnwidth,keepaspectratio]{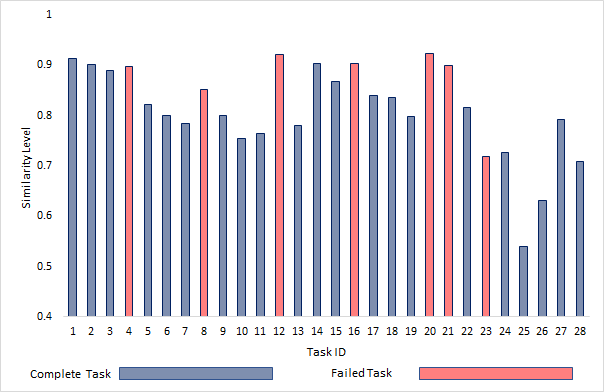}
\caption{Overview of Tasks' Status and Similarity Level in the Platform }
\label{status}
\end{figure}

\begin{figure}[ht!]
\includegraphics[width=1\columnwidth,keepaspectratio]{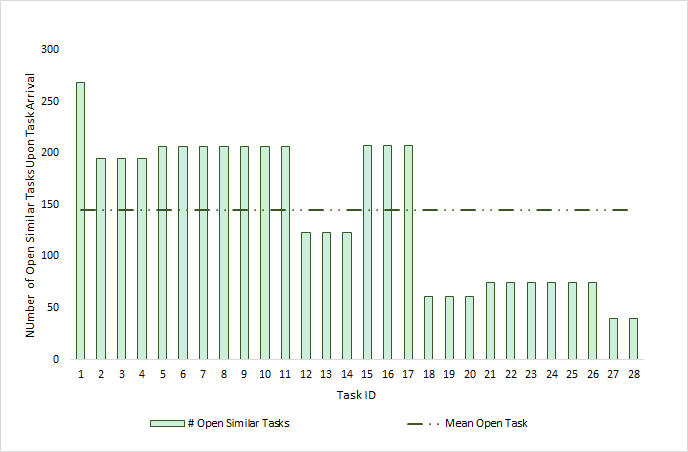}
\caption{Number of Open Tasks in Platform upon Task Arrival}
\label{open}
\end{figure}

If we ignore the task failed based on client request and failed requirements, 28 tasks remain, (see figure\ref{status}). 
Deeper analysis reveals that most of the failed tasks entered the task pool with a similarity above 80\% when compared with the available tasks. 
Also, as figure \ref{open} illustrates, each task competes with an average of 145 similar open tasks upon arrival. The number of open tasks can directly impact the ability of a task to attract suitable workers and lead to task failure. 
It is reported that the degree of task similarity in the pool of tasks directly impacts the task competition level to attract registrants and task success \cite{10.1007/978-3-030-50017-7_7}.

It seems task failure is a result of task competition level in the platform. This observation  motivates us to investigate more and provide a task scheduling recommendation model which helps reduce the probability of task failure in the platform.

\section{\uppercase{Related Work}}

\subsection{Task Scheduling in Crowdsourcing}
Different characteristics of machine and human behavior create delays in product release\cite{ruhe2005art}. This phenomenon leads to a lack of systematic processes to balance the delivery of features with the available resources \cite{ruhe2005art}.
Therefore, improper scheduling would result in task starvation \cite{faradani2011s}. Parallelism in scheduling is a great method to create the chance of utilizing a greater pool of workers \cite{ngo2008optimized,saremi2015empirical}. Parallelism encourages workers to specialize and complete tasks in a shorter period. The method also promotes solutions that benefit the requester and can help researchers to clearly understand how workers decide to compete on a task and analyze the crowd workers performance \cite{faradani2011s}. Shorter schedule planning can be one of the most notable advantages of using CSD for managers \cite{lakhani2010topcoder}.

Batching tasks in similar groups is another effective method to reduce the complexity of tasks and it can dramatically reduce costs\cite{marcus2011human}. Batching crowdsourcing tasks would lead to a faster result than approaches which keep workers separate\cite{bernstein2011crowds}. There is a theoretical minimum batch size for every project according to the principles of product development flow \cite{reinertsen2009principles}. To some extent, the success of software crowdsourcing is associated with reduced batch size in small tasks.
Besides, the delay scheduling method \cite{zaharia2010delay} was specially designed for crowdsourced projects to maximize the probability that a worker receives tasks from the same batch of tasks they were performing. An extension of this idea is introduced a new method called “fair sharing schedule” \cite{ghodsi2011dominant}. In this method, various resources would be shared among all tasks with different demands to ensure that all tasks would receive the same amount of resources. For example, this method was used in Hadoop Yarn. Later, Weighted Fair Sharing (WFS) \cite{difallah2016scheduling} was presented as a method to schedule batches based on their priority. Tasks with higher priority are to be introduced first.

Another proposed crowd scheduling method is based on the quality of service (QOS) \cite{khazankin2011qos}. This is a skill-based scheduling method with the purpose of minimizing scheduling while maximizing quality by assigning the task to the most available qualified worker. This scheme was created by extending standards of Web Service Level Agreement (WSLA) \cite{ludwig2003web}. The third available method method is HIT-Bundle \cite{difallah2016scheduling}. HIT-Bundle is a batch container which schedules heterogeneous tasks into the platform from different batches. This method makes for a higher positive outcome by applying different scheduling strategies at the same time. The method was most recently applied in helping crowdsourcing-based service providers meet completion time SLAs \cite{hirth2019task}. The system works by recording the oldest task waiting time and running a stimulative evaluation to recommend the best scheduling strategy for reducing the task failure ratio.

\subsection{Task Similarity in Crowdsourcing}
Generally, workers tend to optimize their personal utility factor when registering for a task \cite{faradani2011s}. It is reported that workers are more interested in working on similar tasks in terms of monetary prize \cite{yang2015award}, context and technology \cite{difallah2016scheduling}, and complexity level. Context switch generates a reduction in workers’ efficiency \cite{difallah2016scheduling}. However, workers usually try to register for a greater number of tasks than they can complete \cite{yang2016should}. It is reported that a high task similarity level negatively impacts task competition level and team elasticity \cite{10.1007/978-3-030-50017-7_7}. A combination of these observations led to task failure due to: 1) receiving zero registrations for a task based on a low degree of similar tasks and a lack of available skillful workers \cite{yang2015award}, and 2) receiving non-qualified submissions or zero submissions based on a lack of time to work on all the tasks registered by the worker\cite{archak2010money}.

\begin{figure*}[ht!]
\centering
\includegraphics[width=1.0\textwidth]{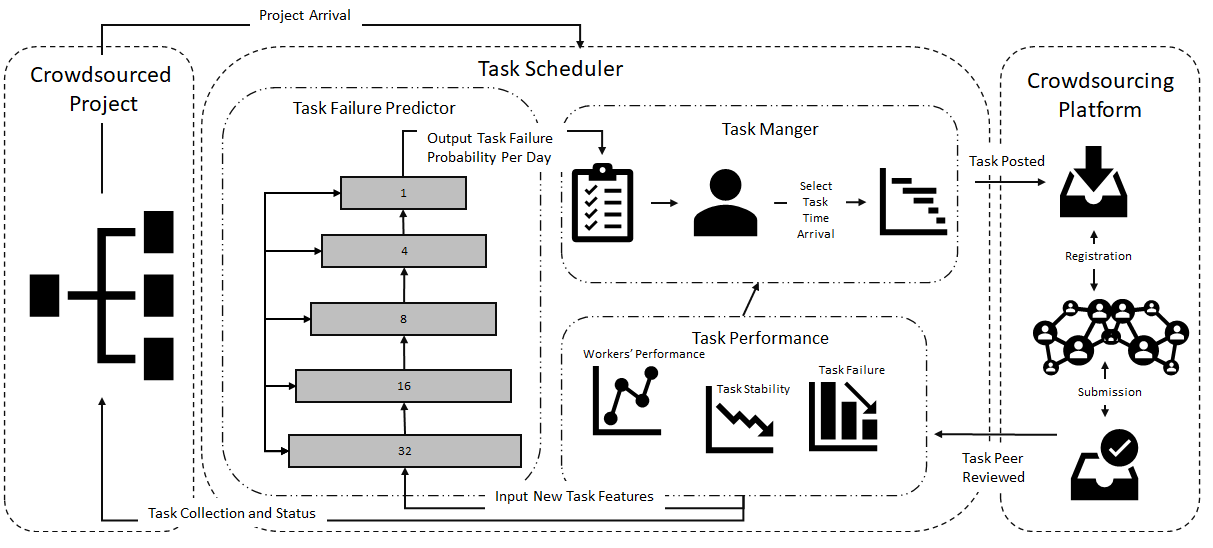}
\caption{Overview of Scheduling Architecture}
\label{diagram}
\end{figure*}

\subsection{Challenges in Crowdsourcing}
Considering the highest rate for task completion and submission acceptance, software managers will be more concerned about the risks of adopting crowdsourcing. Therefore, there is a need for a better decision-making system to analyze and control the risks of insufficient competition and poor submissions due to the attraction of untrustworthy workers. A traditional method of addressing this problem in the software industry is task scheduling. Scheduling is helpful in prioritizing access to resources. It can help managers optimize task execution in the platform to attract the most reliable and trustworthy workers. Normally, in traditional methods, task requirements and phases are fixed, while cost and time are flexible. In a time-boxed system, time and cost are fixed, while task requirements and phases are flexible \cite{cooper2016agile}. However, in CSD all three variables are flexible. This characteristic creates a huge advantage in crowdsourcing software projects.

Generally, improper scheduling could lead to task starvation \cite{faradani2011s}, since workers with greater abilities tend to compete with low skilled workers \cite{archak2010money}. In this case, users are more likely to choose tasks with fewer competitors \cite{yang2008crowdsourcing}. Also, workers who intentionally choose to participate in less popular tasks could potentially enhance winning probabilities, even if workers share similar expertise. It brings some severe problems in the crowd workers trust system to continue performing on a task and causes a lot of dropped and non-completed tasks.  Moreover, tasks with relatively lower monetary prizes have a high probability of registration and completion, which results in only 30\% of problems in the platform being solved \cite{rapoport1966game}. Lower priced tasks may attract higher numbers of workers to compete and consequently increases the chance of starvation for more expensive tasks and project failure. 

The above issues indicate the importance of task scheduling in the platform in order to attract the right amount of trustworthy workers and expertise that will result in a shortened task release time.

\section{\uppercase{Research Design and Methodology}}
\label{sec:Research Design and Methodology}
To solve the scheduling problem, we designed a model to predict the probability of task failure and recommend arrival date based on comparing predicted task failure probabilities. We utilized a neural network model to predict the probability of task failure per day. Then we add a search-based optimizer to recommend arrival day with lowest failure probability. 
This architecture can be operated on any crowdsourcing platform; however, we focused on TopCoder as the target platform. In this method, task arrival date is suggested based on the degree of task similarity in the platform and the reliability of available workers to make a valid submission. Figure \ref{diagram} presents the overview of the task scheduling architecture. Each task is uploaded in the \textit{task scheduler}. The \textit{Task failure predictor} analyzes the probability of failure of an arriving task in the platform based on the number of similar tasks available that day, average similarity, task duration, and associated monetary prize. Then the model recommends a probability of task failure for the assigned date with two days surplus. In next step, the \textit{task manager} selects the most suitable arrival date among the three recommended days and schedules the task to be posted. The result of task performance in the platform is to be collected and reported to the client along with the input used to recommend the posting date.

\begin{table*}[!ht]
\caption{Summary of Metrics Definition} 
\centering 
\begin{tabular}{p{1.5cm} p{4.5cm} p{8cm}}
\hline
Type & Metrics & Definition \\ 
\hline
 & Task registration start date (TR)	& The first day of task arrival in the platform and when workers can start registering for it. Range: (0, $\infty$)) \\
 & Task submission end date (TS)	& Deadline by which all workers who registered for task have to submit their final results. Range: (0, $\infty$)). \\

Tasks attributes & Task registration end date (TRE) & The last day that a task is available to be registered for. Range: (0, $\infty$)). \\
& Monetary Prize (P) & Monetary prize (USD) offered for completing the task and is found in task description. Range: (0, $\infty$)).\\
& Technology (Tech) & Required programming language to perform the task. Range: (0, \#Tech)) \\ 
& Platforms (PLT) & Number of platforms used in task. Range: (0, $\infty$)). \\
\hline
&Task Status & Completed or failed tasks \\
& \# Registration (R)  & Number of registrants that sign up to compete in completing a task before registration deadline. Range: (0, $\infty$). \\
Tasks performance & \# Submissions (S) & Number of submissions that a task receives before submission deadline. Range: (0, \# registrants]. \\
& \# Valid Submissions (VS) & Number of submissions that a task receives by its submission deadline that have passed the peer review. Range: (0, \# registrants]. \\
\hline
\label{metrics}
\end{tabular}
\end{table*}

\subsection{Dataset}

The gathered data set contains 403 individual projects including 4,908 component development tasks and 8,108 workers from Jan 2014 to Feb 2015, extracted from Topcoder website. 
Tasks are uploaded as competitions in the platform and Crowd software workers register for and complete the challenges. On average, most of the tasks have a life cycle of 14 days from the first day of registration to the submission’s deadline. When a worker submits their final files, their submission is reviewed by experts to and labeled as a valid or invalid submission.  
Table\ref{metrics} summarizes the task features available in the data set.

\subsection{Input to the Task Scheduler}
It is reported that task monetary prize and task duration \cite{faradani2011s}\cite{yang2015award}\cite{saremi2020much} are the most important factors in raising competition level for a task. In this research, we are adding the variables considered in our observations (i.e number of open tasks and average task similarity) to the reported list of important factors as input of the presented model.
To help in understanding of the qualities of the task scheduling tool, the input variables of the model, including average task similarity, task duration, task monetary prize, and number of open tasks are defined below. A definition for the probability of task failure in the platform, used as the reward function to train the neural network model, can also be found below.

First we need to understand the degree of task similarity among a set of simultaneously open tasks in the platform.
\textit{Def.1}: Task Similarity ($ {Sim_{i,j}} $) is the similarity between two tasks $ {T_{i}} $ and $ {T_{j}} $ is defined  as the weighted sum of all local similarities across the features listed in Table \ref{Sim} :

\begin{table*}[!ht]
\caption{Features used to measure task distance} 
\centering 
\begin{tabular}{p{6cm} p{8cm}}
\hline
Feature & Description of distance measure $ {Dist_{i}} $ \\ 
\hline
Task Monetary Prize (P) & (${Prize_{i}}$ - ${Prize_{j}}$ ) = ${Prize_{Max}}$ \\
Task registration start date (TR)	& (${TR_{i}}$ - ${TR_{j}}$) = ${DiffTR_{Max}}$ \\
Task submission end date (TS)	& ${TS_{i}}$ - ${TS_{j}}$) = ${Diff_TS_{Max}}$ \\
Task Type & (${Type_{i}}$ == ${Type_{j}}$) ? 1 : 0 \\
Technology (Tech) & Match(${Tech_{i}}$:${Tech_{j}}$)=${NumberOfTechs_{Max}}$ \\ 
Platform (PL) & (${PL_{i}}$ == ${PL_{j}}$) ? 1 : 0 \\
Detailed Requirement & (${Req_{i}} * {Req_{j}})/(|{Req_{i}}|*|{Req_{i}}|$) \\

\hline
\label{Sim}
\end{tabular}
\end{table*}

\[
{Sim_{i,j}} = {{W_{1} * Dist_{1}({T_{i},T_{j}})} + ... +{W_{n} * Dist_{n}({T_{i},T_{j}})}}
\]

\textit{Def.2}: Task Duration ($ {D_{i}} $) is the total available time from task (i) registration start date ($ {TR_{i}} $) to submissions end date ($ {TS_{i}} $):

\[
{D_{i}} = {\sum_{i=0}^{n}{TS_{i} - TR_{i}}}
\]

\textit{Def.3}: Actual Prize ($ {P_{i}} $) is the summation of the  prize that the winner ($ {PW_{i}} $) and runner up( $ {PR_{i}} $) will receive after passing peer review. 

\[
{P_{i}} = {\sum_{i=0}^{n}{PW_{i} + PR_{i}}}
\]

\textit{Def.4}: Number of Open Tasks per day ($ {NOT_{d}} $) is the Number of tasks ($ {T_{j}} $) that are open for registration when a new task ($ {T_{i}} $) arrives on the platform. 

\[
{NOT_{d}} = {\sum_{j=0}^{n}{T_{j}}}
\]
$${where, {TRE_{j} >= TR_{i}} }$$

\textit{Def.5}: Average Task Similarity per day ($ {ATS_{d}} $) is the average similarity score ($ {Sim_{i,j}} $) between the new arriving task ($ {T_{i}} $) and currently open tasks ($ {T_{j}} $) on the platform.  

\[
{ATS_{d}} = {\frac{\sum_{i,j=0}^{n}{Sim_{i,j}}}{NOT_{d}}}
\]
$${where, {TRE_{j} >= TR_{i}} } $$

\textit{Def.6}: Task Failure Rate per day, ($ {TF_{d}} $) is the probability that a new arriving  task ($ {T_{i}} $) does not receive a valid submission and fails given its arrival date.

\[
{TF_{d}} =  1 - {\frac{\sum_{i=0}^{n}{VS_{i}}}{NOT_{i}}}
\]
$${where, {TRE_{j} >= TR_{i}} } $$

\subsection{Output of the Task Scheduler}

The goal of the proposed model is not only to make sure that we can predict the probability of failure for a new arriving task given the arrival date, but also recommend the most suitable posting day to decrease the task failure rate with the surplus of two days. To determine the most optimal arrival date, we run the model and evaluate the result for \textit{arrival day}, \textit{one day after}, and \textit{two days after}.
To predict the probability of task failure in future days, we need to determine the number of expected arriving tasks and associated task similarity scores compared to the open tasks in the future. 

\textit{Def.7}: Rate of Task Arrival per day ($ {TA_{d}} $), Considering that the registration duration (difference between opening and closing dates) for each task is known at any given point in time, the rate of task arrival per day is defined as the ratio of the number of open tasks per day $ {NOT_{d}} $ over the total duration of open tasks per day $ {D_{d}} $.

\[
{TA_{d}} = {\frac{NOT_{d}}{\sum_{j=0}^{n}{D_{j}}}}
\]


By knowing the rate of task arrival per day, the number of open tasks for future days can be determined.

\textit{Def.8}: Number of Open Tasks in the Future $ {OT_{fut}} $ is the number of tasks that are still open given a future date $ {NOT_{fut}} $, in addition to the rate of task arrival per day $ {TA_{d}} $ multiplied by the number of days into the future $ {\Delta}{T_{days}} $.

\[
{OT_{tm}} = {{NOT_{tm}}+{{TA_{d}}}*{\Delta}{T_{days}}}
\]

Also there is a need to know the average task similarity in future days. 

\textit{Def.9}: Average Task Similarity in the Future $ {ATS_{fut}} $, is defined as the number of tasks that are still open given a future date $ {NOT_{fut}} $ multiplied by the average task similarity of this group of tasks $ {ATS_{fut}} $, the average task similarity of the current day $ {ATS_{d}} $ multiplied by the rate of task arrival per day $ {TA_{d}} $ and the the number of days into the future $ {\Delta}{T_{days}} $.

\[
{ATS_{fut}} = {NOT_{fut}}*{ATS_{fut}}+{ATS_{d}}*{TA_{d}}*{\Delta}{T_{days}}
\]

\subsection{Task Failure Predictor}
A fully connected feed forward neural network was trained to predict task failure probability based on the four features described above. The network is configured with five layers of size 32, 16, 8, 4, 2, and 1. 
Training was implemented in in batch sizes of 8 for 50 epochs and the mean-squared error loss function was used. 
 Before running the model, the data set was split into a train/test group and validation group. The train/test group was 80\% of the data set and the validation group was 20\%. We applied  a  K-fold (K=10)  cross-validation method  on  the train/test group to train  the prediction of task  failure probability in the neural network model. For each fold, the 10\% testing portion of the train/test group was cycled and the remaining 90\% was used as train data. We used early stopping to avoid over fitting.  The trained model provided a loss equal to 0.04 with standard deviation of 0.002. 
 
The task manager then uses the output of the task failure predictor to recommend the task arrival day with the minimum failure probability for the schedule plan.
Applying the presented model on the full data set from data set introduced in section IV-A, i.e. all 4908 tasks, yielded a reduction of 8\% ( i.e 75\% to 67\%) in the probability of task failure in the platform.

\section{\uppercase{Case Study and Model Evaluation}}
\label{sec:Case Study and Model Evaluation}

To study the applicability of the proposed method in the real world, we used the system to reschedule the project from the motivating example. (This example is a small part of the full data set; results on the full data set are presented above.) The reschedule result is discussed below: 

\subsection{Result of the Task Failure Predictor}

\begin{figure}[ht!]
\includegraphics[width=1\columnwidth,keepaspectratio]{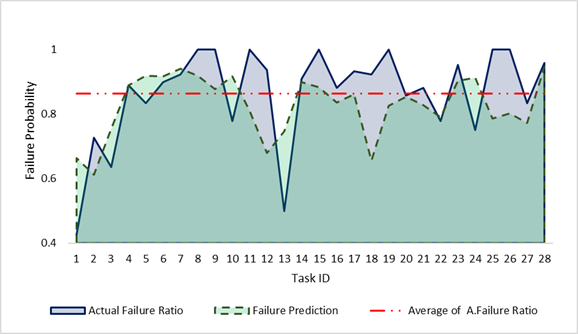}
\caption{Comparison of initial Task Failure Prediction and Actual Failure}
\label{result}
\end{figure}

After testing, the data from the motivation example was provided to the system for evaluation with the goal of figuring out the arrival day with the lowest failure probability per task. Figure \ref{result} presents the initial results of the model. 

As shown in figure \ref{result}, the presented model's arrival date recommendations for the tasks in the sample provide an average failure prediction of 0.83, which is 0.03 lower than the actual scheduling failure.  The result of  initial failure probability prediction by the model in closer to the mean of actual failure, with standard deviation of 0.09. The duration of the project extended an extra day under the new scheduling recommendation, while the probability of task failure was reduced by almost 4\%.
Table \ref{Fstat} summarized the statistics of actual failure and prediction failure for the project.

\begin{table}[!ht]
\caption{summary of actual and predicted failure probabilities of project} 
\centering 
\begin{tabular}{p{2cm} p{2cm} p{2cm}}
\hline
Statistics & Actual P(failure)  & Predicted P(failure)  \\ 
\hline
Min & 0.42 & 0.61 \\
Max & 1	& 0.94 \\
Mean & 0.86	& 0.83 \\
Median & 0.90 & 0.84 \\
Std & 0.15 & 0.09 \\
\hline
\label{Fstat}
\end{tabular}
\end{table}

\begin{figure}[ht!]
\includegraphics[width=1\columnwidth,keepaspectratio]{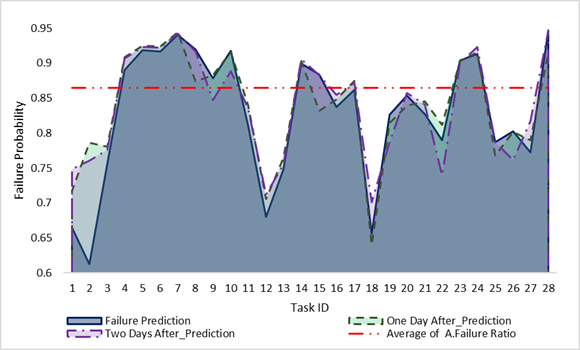}
\caption{Details of Failure Prediction per Task for all level of predictions}
\label{recom}
\end{figure}

In next step, the model provides predictions of task failure probability for one and two days after the actual arrival date of the task. This 
result is used by the task manager to determine if the task should be posted in the future instead of today. Figure \ref{recom} illustrates the result of the failure prediction of all the three dates.

 Tasks 8,15,18,20,25,28 received the lowest prediction of failure probability on the second day with an average prediction of 0.81. Tasks 9, 10, 15, 22, 23, 26 received the lowest prediction of failure probability on day 3 with an average prediction of 0.8. The rest of the tasks received the lowest prediction of failure probability on the first day with an average of 0.81. However, not only was the average of all the three predictions lower than the actual failure prediction, but also most of the prediction points in all three days were lower than the average of actual failure.

\begin{figure}[ht!]
\includegraphics[width=1\columnwidth,keepaspectratio]{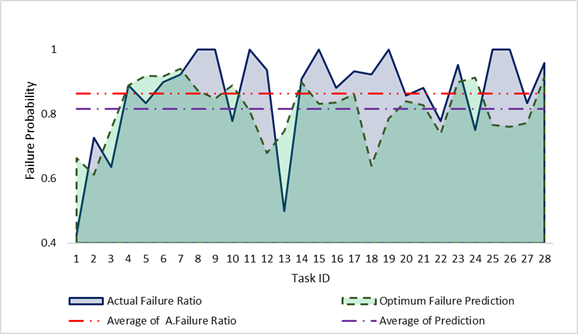}
\caption{Comparison of Probability of Task Failure Prediction for final Schedule and Actual Schedule}
\label{final}
\end{figure}

With access to the 3-day outlook of task failure predictions from the model and the evaluation in figure \ref{recom}, the task manager can more effectively schedule tasks by choosing the minimum task failure probability from the model results for each task. Figure \ref{final} presents the failure probability for the project following the lowest failure prediction per task in comparison with the actual task failure. It is clear that the model's recommended schedule provides a lower and more stationary probability of task failure with an average of 0.81, while the average probability of task failure for the actual task schedule is 0.86. The recommended task scheduling plan provides a minimum failure prediction of 0.61 and maximum of 0.94 with standard deviation of 0.09. The resulting accuracy of the model is 0.896.

\begin{figure}[ht!]
\includegraphics[width=1\columnwidth,keepaspectratio]{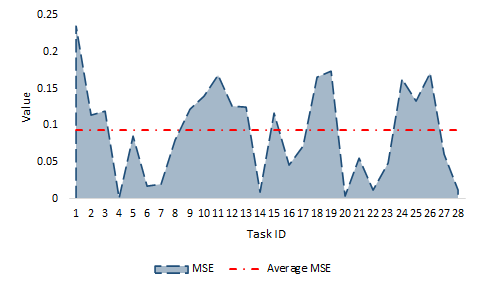}
\caption{MSE for Probability of Task Failure per Task}
\label{MSR}
\end{figure}

\begin{figure*}[ht!]
\centering
\includegraphics[width=1.0\textwidth]{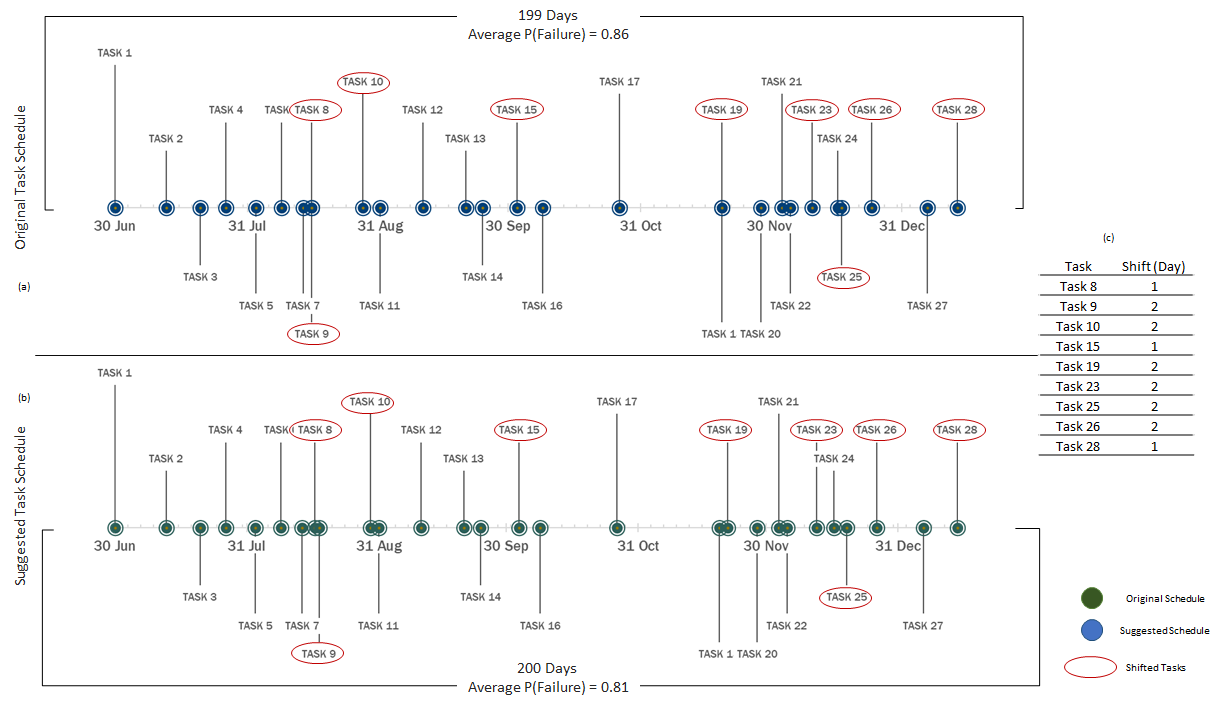}
\caption{Project Timeline }
\label{Finalgant}
\end{figure*}

Figure \ref{Finalgant} illustrates the summary of original task timeline v.s. final task time line. \ref{Finalgant}(a) presents the original project timeline, and \ref{Finalgant}(b) shows the suggested project timeline by the presented model. \ref{Finalgant}(c)represents the duration of each task when shifted to achieve lowest failure probability.

To evaluate the model performance, we applied the Mean Square Error (${MSE}$) metric to estimate the difference between the actual failure probability and the predicted failure probability of the same arrival day according to available data. Figure \ref{MSR} presents the MSE for failure prediction per task. The average MSE is 0.09 with a minimum of 0.001 for task 3 and a maximum of 0.23 for task 1, with a standard deviation of 0.06.

\subsection{Model Evaluation}
To compare the performance of the proposed model, we applied a K-fold (K=10)  cross-validation on the data set to predict the probability of task failure based on four different prediction approaches. The estimated probabilities of task failure are used to compute four popular performance measures that are widely used in current prediction systems for software development: 1- Mean Square Error (${MSE}$), 2- Median of Mean Square Error (${MdMSE}$), 3- Standard Deviation of mean Square Error (${StdMSE}$), 4- Percentage of the estimates with Mean Square Error less than or equal to N\% (${Pred(N)}$).

\begin{figure}[ht!]
\includegraphics[width=1\columnwidth,keepaspectratio]{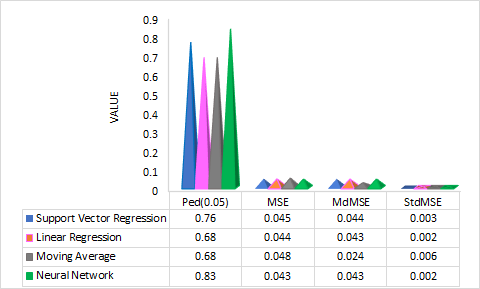}
\caption{Performance of Task Failure Probability by Each Approach}
\label{MSE}
\end{figure}

The primary result of this analysis is shown in figure \ref{MSE}.
It is clear that Neural Network analysis has a better predictive performance according to ${Pred(0.05)}$ and also has almost the lowest error rate with the MSE of 0.043\%. The SVR recreation is the runner up in terms of performance with the MSE of 0.045\%. Interestingly, the moving average and linear regression provided the same level of performance based on ${Pred(0.05)}$, while linear regression provides the lower MSE of 0.044.

\subsection{Threats to Validity}

First, the study only focuses on competitive CSD tasks on the TopCoder platform. Many more platforms do exist, and even though the results achieved are based on a comprehensive set of about 5,000 development tasks, the results cannot be claimed to be externally valid. There is no guarantee the same results would remain exactly the same in other CSD platforms.

Second, there are many different factors that may influence task similarity, task success, and task completion. Our similarity algorithm and task failure probability-focused approach are based on known task attributes in TopCoder. Different similarity algorithms and task failure probability-focused approaches may lead us to different, but similar results.

Third, the result is based on tasks only. Workers' network and communication capabilities are not considered in this research. In the future, we need to add this level of research to the existing one.

\section{\uppercase{Conclusion and future work}}
\label{sec:Conclusion and future work}

CSD provides software organizations access to an infinite, online worker resource supply. Assigning tasks to a pool of unknown workers from all over the globe is challenging. A traditional approach to solving this challenge is task scheduling. Improper task scheduling in CSD may cause zero task registrations, zero task submissions or low qualified submissions due to uncertain worker behavior, and consequentially, task failure. This research presents a new scheduling method based on a neural network. The method reduces the probability of task failure in CSD platforms. The experimental result show a reduction in project failure probability of up to 4\% while maintaining the same project duration.

In future research, we will focus on expanding the model to a more complicated framework that incorporates available worker similarity and considers the impact of the workers' competition performance regarding the task success to further improve efficiency in the scheduling model. Moreover, the presented neural network model will be used as a fitness function in an evolutionary algorithm based scheduling.

\bibliographystyle{apalike}
{\small
\bibliography{Refrence}}



\end{document}